\documentstyle[12pt]{article}
\def\gev{{\rm \, Ge\kern-0.125em V}}
\begin{document}
\begin{titlepage}
\pagestyle{empty}
\baselineskip=21pt
\rightline{McGill 95--42}
\rightline{August 1995}
\vskip .2in
\begin{center}
{\large{\bf Hidden Finite Symmetries in String Theory and}}

{\large{\bf Duality of Dualities}}
\end{center}
\vskip .35in
\begin{center}
Nemanja Kaloper

Department of Physics, McGill University

Montr\'eal, Qu\'ebec, Canada H3A 2T8

{\it email: kaloper@hep.physics.mcgill.ca}

\end{center}
\vskip .45in
\centerline{ {\bf Abstract} }
\baselineskip=18pt
Different compactifications of six-dimensional string theory
on $M_4 \times T^2$ are considered.
Particular attention is given to
the roles of the reduced modes as the $S$ and $T$ fields. It is
shown that there is a discrete group of invariances of an
equilateral triangle hidden in the model. This group is
realized as the interchanges of the two-form fields present in
the intermediate step of dimensional reduction in five dimensions.
The key ingredient for the existence of this group is the presence
of an additional $U(1)$ gauge field in five dimensions, arising
as the dual of the Kalb-Ramond axion field strength.
As a consequence, the theory
contains more four-dimensional $SL(2,R)$ representations, with
the resulting complex scalar
axidilaton related to the components of the Kaluza-Klein vector
fields of the naive dimensional reduction. An immediate byproduct
of this relationship is a triadic correspondence
among the fundamental string, the solitonic string, and a singular
Brinkmann pp wave.

\vskip.5in
\centerline{\it Submitted to Physics Letters {\bf B}}

\end{titlepage}
\baselineskip=18pt
{\newcommand{\la}{\mbox{\raisebox{-.6ex}{~$\stackrel{<}{\sim}$~}}}}
{\newcommand{\ga}{\mbox{\raisebox{-.6ex}{~$\stackrel{>}{\sim}$~}}}}
There has been much interest recently in studying strong-weak
coupling duality symmetry and the relationship of the associated
$S$ and $T$ fields in string theory in four dimensions
\cite{SEN1}-\cite{strom}.
These symmetries assist our taxonomy of different
string constructions by providing links between them, predict
the existence and properties of non-trivial topological
configurations, and give means for constructing new configurations
from the old ones. Generically, two kinds of dualities are
encountered in string theory: 1) the target space ($T$) duality,
which is realized as an $O(d,d)$
transformation group of moduli and gauge fields leaving the
effective action invariant, and 2) the strong/weak coupling
($S$) duality, which appears
as an $SL(2,R)$ group of transformations of the complex scalar
axidilaton
field and gauge fields, that keep the equations of motion
invariant but not the action in the naive dimensional
reduction\footnote{There exists a formulation of the theory where
the manifest invariance of the action under $SL(2,R)$ transformations
is restored, see \cite{sensch}.}. Each group has
a discrete subgroup believed to be an exact
symmetry of string theory both in the string loop expansion
and in the inverse string tension $\alpha'$
expansion, with the difference that $T$ should be perturbatively
exact in $\alpha'$ and nonperturbatively exact in the string loop
expansion, and vice-versa for $S$. It has been shown that in
certain special cases these two symmetries can be interchanged,
establishing a connection between either two different
string theories (e.g. heterotic and type IIa, with specific
compactifications from ten to four dimensions
\cite{dk,duff2,sen2}), or
between different compactifications within the same theory
(e.g. regarding the duality of dualities map as
a relationship between different vacua
in the same string theory,
which is allowed for special solutions with
trivial fermions and Yang-Mills gauge fields)
\cite{bine,dk,duff2}.  In this letter I will
focus on the latter case, and assuming that the model considered
is the heterotic string theory demonstrate how the results
obtained so far can be extended to obtain another, alternative,
compactification providing new
representations for the dualities.
These symmetries become manifest
when the Kalb-Ramond axion
field strength is replaced by its Hodge dual
(weighted with the dilaton), which in
five dimensions plays the role of an additional
vector gauge field \cite{witten}.
The resulting representations are related by a finite group of
symmetries of the equilateral triangle $D_4$, realized by
mixing the modulus with the dilaton and permuting the
resulting $U(1)$ gauge fields
in the intermediate step of dimensional
reduction in five dimensions. One of these transformations
is just the standard scale-factor $T$ duality in five dimensions,
whereas another is precisely the duality of dualities
transformation conjectured by Duff and Khuri \cite{dk}, and
derived in more general circumstances by Duff \cite{duff2}.
As the $D_4$ group has only two independent generators,
we see that these two transformations are in fact
sufficient to describe the full action of the group.  The result
of their combinations, however, is non-trivial because they
lead to different compactifications to four dimensions,
interchanging four dimensional moduli with the axidilaton, and
concomitantly realizing a multiplicity of string dualities.
I will finally discuss a straightforward application to the
fundamental string solution in six dimensions, and using its
dual relationship to the solitonic string solution and
a Brinkmann pp wave, propose that there should be a string
subtraction scheme in which the solitonic solution should
be exact to all orders in the inverse string tension expansion.

In order to define the $T$ and $S$ symmetries, we need to look at
the effective theory in four dimensions. Here only
the basics of dimensional
reduction for the bosonic sector of string theory will be presented
(see e.g. \cite{MS} for a more complete account).
Our starting point is the effective action
describing the graviton multiplet in
six dimensions, to the lowest order in $\alpha'$
\cite{MS,duff2,sen2}:
\begin{equation}
\label{sact}
S = \int d^6x \sqrt{{\bar g}} e^{- \Phi}
\Bigl\{\bar{R} + (\bar{\nabla} \Phi)^2 - \frac{1}{12}
\bar{H}^2_{\mu\nu\lambda}\Bigr\}
\end{equation}
The bar denotes six-dimensional quantities throughout the
article. The fields $\Phi$ and
$\bar{H}_{\mu\nu\lambda} = \partial_\mu
\bar{B}_{\nu\lambda} + cyclic~$
$permutations$ are the six dimensional
dilaton and three-form axion fields respectively.
The explicit representation of the string duality symmetries
is accomplished by the Kaluza-Klein reduction of this theory to
four dimensions, resulting in an
action with additional scalar and gauge fields coupled
to the graviton, axion and dilaton.
The gauge fields arise from the
cross-terms in the metric (``shift functions")
and the two-form axion potential, and are invariant
under the corresponding $U(1)$ symmetries.
The scalars are the breathing modes
of the compact directions and the two-form axion field
components living in the compact submanifold.
The duality symmetries are
included in the model via the couplings of these additional fields.
Thus, in terms of the six-dimensional
background, the metric and matter fields are
\begin{eqnarray}
\label{dans}
d{\bar{s}}^2 &=& {g}_{\mu\nu} dx^{\mu} dx^{\nu} + {G}_{AB}
(dy^A + V^A{}_{\mu} dx^{\mu})
(dy^B + V^B{}_{\nu} dx^{\nu}) \nonumber \\
\bar{B} &=& \frac{1}{2}
\Bigl( {B}_{\mu\nu} - \frac{1}{2}
\bigl( V^A{}_{\mu} {B}_{\nu A} -  V^A{}_{\nu}
{B}_{\mu A} \bigr) +
{B}_{AB} V^A{}_{\mu} V^B{}_{\nu} \Bigr) dx^{\mu}dx^{\nu}
\nonumber \\
&& ~+  \Bigl( {B}_{\mu A} - {B}_{AB} V^B{}_{\mu} \Bigr)
dx^{\mu} dy^A + \frac{1}{2} {B}_{AB} dy^{A}dy^{B} \\
\Phi&=&\phi + \frac{1}{2} \ln |\det({G}_{AB})| \nonumber
\end{eqnarray}
The vector fields $V^{A}{}_{\mu},B_{\mu A}$
are the $U(1)$ gauge fields mentioned above which are
coming from the cross-terms in the metric and
the axion, respectively. Their field strengths will be denoted
by $V^{A}{}_{\mu\nu}$ and $H_{\mu\nu A}$.
The corrections in (\ref{dans})
proportional to $V$ are necessary to
disentangle the resulting
four-dimensional gauge symmetries.

After straightforward but tedious algebraic
manipulations with the reduced action,
using gauge-invariance as the guide and switching
between the tangent-space and holomorphic bases, we
can rewrite the action in the following form:
\begin{equation}
\label{fact}
S = \int d^4x \sqrt{{g}} e^{- \phi} \Bigl\{{R}
 + ({\nabla} \phi)^2 + \frac{1}{8} Tr \bigl({\cal L}
 \nabla {\cal M}\bigr)^2
- \frac{1}{4} {\cal F}^T_{\mu\nu} {\cal M} {\cal F}^{\mu\nu}
- \frac{1}{12} {H}^2_{\mu\nu\lambda}\Bigr\}
\end{equation}
The capital $T$ denotes matrix transposition.
The $\sigma$-model fields ${\cal M}$ appear after
rearranging the scalar moduli fields.
The correspondence is given by
\begin{equation}\label{w4}
{\cal M}~=~\pmatrix{
g^{-1}&-g^{-1}b \cr
bg^{-1}&g - b^{T}g^{-1}b \cr} ~~~~~~~~~~
{\cal L}~=~\pmatrix{~0~&{\bf 1} \cr
{\bf 1}&~0~\cr}
\end{equation}
where $~g~$, $~b~$ and ${\bf 1}$ are $~2 \times 2~$ matrices
defined by the dynamical degrees of freedom of the metric and the
axion:$~g~=~\bigl(G_{AB}\bigr)$ and $b~=~\bigl(B_{AB}\bigr)$.
The vector multiplet is obtained from the
Kaluza-Klein and axionic gauge fields
as follows:
\begin{equation}\label{vectm}
{\cal A}_{\mu}~=~\pmatrix{ B_{\mu A}\cr
                            V^A{}_{\mu} \cr}
{}~~~~~~~~~~
{\cal F}_{\mu\nu}~=~\pmatrix{ H_{\mu\nu A}\cr
                               V^A{}_{\mu\nu} \cr}
\end{equation}
and the axion field strength can be rewritten as
\begin{equation} \label{axfsin}
 {H}_{\mu\nu\lambda} = \partial_{\mu} {B}_{\nu\lambda}
- \frac{1}{2}
{\cal A}^T_{\mu} {\cal L} {\cal F}_{\nu\lambda} +
cyclic ~permutations
\end{equation}
Note
that ${\cal M}^{T} = {\cal M}$ and
${\cal M}^{-1} = {\cal L}{\cal M}{\cal L}$. Thus we see
that ${\cal M}$ is a symmetric element of $O(2,2,R)$. Therefore an
$O(2,2,R)$ rotation ${\cal M} \rightarrow \Omega {\cal M} \Omega^{T}$
and ${\cal F} \rightarrow \Omega {\cal F}$, while
changing ${\cal M}$ and ${\cal F}$, is a symmetry
of the action and the equations of motion.
The continuous $O(2,2,R)$ symmetry is believed to be broken down
to $O(2,2,Z)$ by non-perturbative effects. This symmetry contains the
$T$-duality symmetry of string theory \cite{SEN1,MS}.

To see how the other duality symmetry, $S$, comes into
play, we need to first conformally rescale
the metric to the Einstein frame, which in four dimensions
is given by $\hat g_{\mu\nu} = \exp(-\phi) g_{\mu\nu}$, and
then replace the three-form axion field strength in
four dimensions by its dual pseudoscalar
field. The correspondence between them is
\begin{equation}
\label{axiondef}
H_{\mu\nu\lambda} = e^{2 \phi} \sqrt{\hat g}
\epsilon_{\mu\nu\lambda\rho} \hat \nabla^{\rho} a
\end{equation}
The hat here denotes the Einstein conformal frame.
If we introduce the complex axidilaton field
$\Psi = a + i \exp(-\phi)$ and the
self-dual and antiself-dual parts of the gauge fields
$\hat {\cal F}^{\pm}_{\mu\nu} =
\hat {\cal F}_{\mu\nu} \pm i {\cal L} {\cal M}{^{*}
\hat {\cal F}_{\mu\nu}}$,
where
${^{*}\hat{\cal F}_{\mu\nu}} = \frac{1}{2}  \sqrt{\hat g}
\epsilon_{\mu\nu\lambda\rho} \hat {\cal F}^{\lambda\rho}$
is the Hodge dual of the two-form $\hat {\cal  F}_{\mu\nu}$,
we can rewrite the action (\ref{fact}) as
\begin{equation}
\label{axidilacti}
S = \int d^4x \sqrt{\hat g} \Bigl\{ \hat R
+ \frac{2 \hat \nabla_{\mu} \Psi \hat \nabla^{\mu}
 \Psi^{\dagger}}{(\Psi - \Psi^{\dagger})^2}
+ \frac{1}{8} Tr \bigl({\cal L} \hat \nabla {\cal M}\bigr)^2
+ \frac{i}{16} \Psi \hat {\cal F}^{-T}_{\mu\nu} {\cal M}
\hat{\cal F}^{-\mu\nu}
- \frac{i}{16} \Psi^{\dagger} \hat{\cal F}^{+T}_{\mu\nu}
{\cal M} \hat{\cal F}^{+\mu\nu} \Bigr\}
\end{equation}
where $\dagger$ denotes complex conjugation.
A careful examination of the equations of motion
derived from this action shows that they are
invariant under the following set of transformations:
\begin{eqnarray}
\label{sl2r}
\Psi \rightarrow \Psi' =
\frac{\alpha \Psi + \beta}{\gamma \Psi + \delta} ~~~~~~
\hat {\cal F}^{-}_{\mu\nu} \rightarrow
\hat {\cal F}'^{-}_{\mu\nu} = (\gamma \Psi + \delta)
\hat {\cal F}^{-}_{\mu\nu}
{}~~~~~ \alpha \delta - \beta \gamma =1
\end{eqnarray}
where $\alpha$, $\beta$, $\gamma$ and $\delta$ are all real
numbers. Also, $\hat {\cal F}^{+}_{\mu\nu}$ transforms like
the complex conjugate of  $\hat {\cal F}^{-}_{\mu\nu}$.
The equations of motion remain invariant under
(\ref{sl2r}) because the gauge equations of motion
(Euler-Lagrange and Bianchi)
are interchanged and the axidilaton
equation itself is invariant, as well as the gauge
energy-momentum tensor
(the term from which it is derived changes
by a boundary term only). However, the action
(\ref{axidilacti}) itself is not invariant.
This can be seen from the fact that (\ref{sl2r}) changes
the sign of the gauge terms in the action.
The transformations (\ref{sl2r}) combine to form the SL(2,R) group.
This symmetry group is referred to as the
strong/weak, or $S$, duality. Again, it
is believed to be broken down to $SL(2,Z)$ by instanton
effects; the relevant physical
symmetry is thus this discrete group \cite{SEN1,sensch}.

The above illustrates the essential properties of the two
duality symmetries. These two symmetries appear to be
considerably different
in that $T$ leaves the action invariant and $S$ does not.
This problem is easily resolved with the
introduction of several Lagrange multipliers,
after which the action can be rewritten
in a manifestly $O(2,2)$ {\it and} $SL(2)$
invariant form \cite{sensch}. It
is then reasonable to ask if the two groups can be related or perhaps
interchanged. This ``duality of dualities" has in fact been recently
shown to hold between different string theories
\cite{duff2,sen2}, and the implications
are still being studied. Essentially, it is based on
the possibility to reduce the model
after dualizing the three-form in six dimensions,
which transforms the four-dimensional
moduli fields, as will be outlined later.

A generalization of this duality of dualities can be attained in the
following way. The dimensional descent can be taken more gradually.
Choosing any one of the two cyclic coordinates,
the starting six-dimensional action can be first
reduced to five dimensions, again according to
the standard Kaluza-Klein prescription.
Because the axion moduli are absent, the formulas for the
reduced modes are now simpler than in the four-dimensional
case. The reduced axionic contributions and the dilaton
are given by
\begin{eqnarray}
\label{kk}
d\bar s^2_6 &=& g_{5\mu\nu} dx^\mu dx^\nu + \zeta e^{2 \sigma}
\bigl(dy + V_\mu dx^\mu \bigr)^2 \nonumber \\
\bar B &=& \frac{1}{2} \Bigl( {B}_{5\mu\nu} - \frac{1}{2}
\bigl( V_{\mu} {B}_{\nu} -  V_{\nu} {B}_{\mu} \bigr)\Bigr)
 dx^\mu \wedge dx^\nu + B_{\mu} dx^\mu \wedge dy \\
\Phi &=& \phi_5  + \sigma \nonumber
\end{eqnarray}
For the sake of simplicity, we have dropped the superscript $``y"$
from the gauge fields $V^y{}_{\mu}$ and ${B}_{\mu y}$, but
have retained the subscript $``5"$ on the metric, the dilaton, and
the two-form axion to
distinguish them from the four-dimensional quantities.
In this equation $\zeta = \pm 1$ allows for the possibility that
the reduced coordinate $y$ is either space- or
time-like, respectively.
(We leave this option open because the duality
transformations can be used for generating solutions, and
so are insensitive of the topological character of the Killing
coordinates.) One must keep in mind that the gauge fields $V$
and $B$ in five dimensions are {\it not} identical to the
gauge fields appearing in the direct $6 \rightarrow 4$ reduction.
Rather, they contain different admixtures of the
effective four-dimensional gauge fields. Obviously, since we could
choose either of the two compact coordinates
to integrate out first, there will be in
general two different possibilities for the intermediate
five-dimensional theory.
The relationship between the five- and four-dimensional fields
can be found by comparing (\ref{dans}) and (\ref{kk}), and will
not be listed here. Then, the reduced
three-form axion in five dimensions is
\begin{equation}
\label{rda}
{H}_{5\mu\nu\lambda} = \partial_{\mu} {B}_{5\nu\lambda} - \frac{1}{2}
V_{\mu} {H}_{\nu\lambda} - \frac{1}{2} {B}_{\mu} V_{\nu\lambda} +
cyclic ~permutations
\end{equation}
The reduced action then becomes, after dividing by $\int dy$,
\begin{equation}
\label{rsact}
S = \int d^5x \sqrt{g_5} e^{- \phi_5} \Bigl\{ R_5
 + ({\nabla} \phi_5)^2  - ({\nabla} \sigma)^2
- \frac{\zeta}{4} e^{2\sigma}{V}^2_{\mu\nu}
- \frac{\zeta}{4} e^{-2\sigma}{H}^2_{\mu\nu}
- \frac{1}{12}{H}^2_{5\mu\nu\lambda}\Bigr\}
\end{equation}
At this point, we need to conformally transform the metric
to the five-dimensional Einstein frame. This is necessary in
order to bring the gauge-scalar couplings in
the manifestly symmetric form, as we will show shortly.
Namely, after we conformally transform to the Eisntein
frame, and rescale the dilaton field $\phi_5$ to
$\eta = \phi_5/\sqrt{3}$, we will find that
the scalar fields $\vec \sigma = (\sigma, \eta)$
can be viewed as an $O(2)$
doublet. Moreover, the Einstein frame metric will be decoupled, and
thus invariant under these isospin rotations. We will then show
that the gauge
kinetic terms couple to the scalar doublet via a set of
Toda-like functions $\exp(\vec Q_k \cdot \vec \sigma)$
for some constant isovectors $\vec Q_k$, $k \in {1,\ldots,3}$,
to be determined. These couplings can transform
covariantly among themselves under discrete subgroups of $O(2)$,
if the isovectors lie on vertices of regular polygons.
Obviously, if these transformations are
followed by permutations of the associated gauge fields, the
induced representation may be a symmetry of the action.
Thus, the conformal transformation is given by
$\hat g_{5\mu\nu} = \exp(-2\phi_5/3) g_{5\mu\nu}$.
The next step is to replace the
Kalb-Ramond axion field strength by its Hodge
dual two-form, which introduces the third gauge field
in the model \cite{witten}.
This dualization of the Kalb-Ramond field strength
is carried out with the help of a Lagrange multiplier vector
field $X_{\mu}$, to account properly for the axion Bianchi identity
$\epsilon^{\sigma\rho\mu\nu\lambda} \partial_{\rho}
{H}_{5\mu\nu\lambda} =
- \frac{3}{2} \epsilon^{\sigma\rho\mu\nu\lambda}
V_{\rho\mu} {H}_{\nu\lambda}$. This vector field actually becomes
the gauge potential for the gauge field $X_{\mu\nu}$.
The relation between it and the Kalb-Ramond field strength is
\begin{equation}\label{XKRd}
{H}_{5\mu\nu\lambda} =  \frac{1}{2} e^{4 \phi_5/3} \sqrt{\hat g_5}
\epsilon_{\sigma\rho\mu\nu\lambda}
\hat X^{\sigma\rho}
\end{equation}
The full five-dimensional action in the Einstein frame can then be
rewritten as
\begin{eqnarray}\label{E5}
S = \int d^5x \sqrt{\hat g_5}  \Bigl\{ \hat R_5
&-&({\hat \nabla} \eta)^2 -
({\hat \nabla} \sigma)^2
- \frac{\zeta}{4} e^{2(\sigma-\eta/\sqrt{3})}\hat {V}^2_{\mu\nu}
\nonumber \\
&-& \frac{\zeta}{4} e^{-2(\sigma+\eta/\sqrt{3})}\hat {H}^2_{\mu\nu}
- \frac{\zeta}{4} e^{4\eta/\sqrt{3}}\hat {X}^2_{\mu\nu}
+ \frac{1}{4} \frac{\epsilon^{\sigma\rho\mu\nu\lambda}}{\sqrt{\hat g_5}}
X_{\sigma} V_{\rho\mu} {H}_{\nu\lambda}
\Bigr\}
\end{eqnarray}
We note that the anomaly-like term, trilinear in the gauge fields, is
symmetric under arbitrary permutations of these fields.
To extend the permutation symmetry of the anomaly-like term,
we note that in light of the aforementioned $O(2)$ isospace
interpretation, the coupling vectors
which determine the scalar-gauge couplings are given as
$\vec Q_1 =(2,-2/\sqrt{3})$, $\vec Q_2 = (-2,-2/\sqrt{3})$,
and $\vec Q_3 = (0,4/\sqrt{3})$ and
lie on the vertices of an equilateral triangle. Since the
invariance transformations of the coupling triangle are
in fact a discrete subgroup of $O(2)$, we see that they
must be invariances of the action. Indeed, it is straightforward to
verify that the transformations
from the group $D_4$,
of the form
\begin{equation}\label{D41}
\pmatrix{\sigma\cr
         \frac{\phi_5}{\sqrt{3}} \cr}
\rightarrow
\pmatrix{\sigma'\cr
         \frac{\phi'_5}{\sqrt{3}} \cr} = \bar \Omega_2
\pmatrix{\sigma\cr
         \frac{\phi_5}{\sqrt{3}} \cr}
{}~~~~~~~
\pmatrix{V_{\mu}\cr
         B_{\mu}\cr
         X_{\mu}\cr}
\rightarrow
\pmatrix{V'_{\mu}\cr
         B'_{\mu}\cr
         X'_{\mu}\cr} = \bar \Omega_3
\pmatrix{V_{\mu}\cr
         B_{\mu}\cr
         X_{\mu}\cr}
\end{equation}
where the matrices $\bar \Omega_2$ and
$\bar \Omega_3$ belong to a two-dimensional
and a three-dimensional representation of $D_4$,
respectively, leave the form of the action unchanged.
Under these transformations, the three different gauge
fields of the five-dimensional action are interchanged
together with the couplings. We note that because the
five-dimensional Einstein frame metric does not change
under (\ref{D41}),
the world-sheet metric transforms according to
$g_{5\mu\nu} \rightarrow g'_{5\mu\nu} =
\exp \Bigl( 2(\phi'_5 - \phi_5)/3 \Bigr) g_{5\mu\nu}$.

One of these transformations is just the standard scale-factor
$T$ duality in five dimensions, which interchanges
$V$ and $B$, and inverts the $\sigma$ field: $V \leftrightarrow B$ and
$\sigma \rightarrow - \sigma$. Another transformation, with
$\sigma'=(\sigma-\phi_5)/2$, $\phi'_5=-(3\sigma+\phi_5)/2$, and
$B \leftrightarrow X$
is precisely the duality of dualities transformation
derived by Duff \cite{duff2}.
This can be seen as follows. The original duality of dualities
map was realized in six dimensions, starting with the action
(\ref{sact}), and
transforming the fields according to \cite{duff2,sen2}
\begin{eqnarray}\label{Duff2}
\bar g_{\mu\nu} \rightarrow \bar g'_{\mu\nu}
&=& e^{-\Phi} \bar g_{\mu\nu} \nonumber \\
\bar H_{\mu\nu\lambda} \rightarrow \bar H'_{\mu\nu\lambda}
&=& \frac{1}{6} e^{-\Phi} \sqrt{\bar g}
\epsilon_{\mu\nu\lambda\alpha\beta\gamma}
\bar H^{\alpha\beta\gamma} \\
\Phi \rightarrow \Phi' &=& - \Phi \nonumber
\end{eqnarray}
The action in terms of the transformed fields is:
\begin{equation}\label{sactD}
S = \int d^6x \sqrt{{\bar g'}} e^{- \bar\Phi'} \Bigl\{ \bar{R'}
 + ({\nabla} \bar\Phi')^2 - \frac{1}{12}
\bar{H'}^2_{\mu\nu\lambda}\Bigr\}
\end{equation}
i.e., it is form-invariant.
If we now attempt to represent these formulas in five
dimensions, using (\ref{kk}),
we immediately obtain these
transformation rules for the reduced fields:
\begin{eqnarray}\label{Ddred}
2\sigma' &=&  \sigma -\phi_5  \nonumber \\
2\phi'_5 &=& -\phi_5 -3\sigma \nonumber \\
g'_{5\mu\nu} &=& e^{2(\phi'_5 - \phi_5)/3} g_{5\mu\nu} \\
V'_{\mu} &=&  V_{\mu} \nonumber
\end{eqnarray}
which we recognize as a part of the sought correspondence.
The last step is to find how the reduced axion fields transform.
In order to do it, we first have to separate between
the reduced two- and three-forms. Noting from (\ref{kk})
that $\bar H_{\mu\nu y} = H_{\mu\nu}$, and also recalling
the definition of the reduced two-form axion
potential (\ref{kk}),
we find that
\begin{eqnarray}\label{Ddaxred2}
H'_{\mu\nu} &=& - \frac{\zeta}{6} e^{-\phi_5} \sqrt{g_5}
\epsilon_{\mu\nu\alpha\beta\gamma}
H_5^{\alpha\beta\gamma} \nonumber \\
\bar H'_{\mu\nu\lambda} &=& \frac{1}{2} e^{-\phi_5} \sqrt{g_5}
\epsilon_{\mu\nu\lambda\alpha\beta} \bigl( e^{-2\sigma}
H^{\alpha\beta} - \zeta V_{\gamma} H_5^{\alpha\beta\gamma} \bigr)
\end{eqnarray}
Note that (\ref{Ddaxred2}) are given in terms of the
string-frame metric. It is obvious that RHS of
the first of these equations is just the inverse
of the Eq. (\ref{XKRd}). This gives
$H'_{\mu\nu} = X_{\mu\nu}$. Also, in the
second equation, $\zeta$ does not multiply
the first term in the parenthesis
because it appeared through the square $\zeta^2=1$.
Now, we must recall that $\bar H_{\mu\nu\lambda}$
is not the gauge invariant quantity in five dimensions.
It transforms anomalously under
Kaluza-Klein gauge transformations. The second term in
the second equation of (\ref{Ddaxred2}) accounts for this.
We rectify this problem by replacing it with
the gauge-invariant field strength, which we find by
combining (\ref{kk}) and (\ref{rda}):
\begin{equation}\label{gifs}
H_{5\mu\nu\lambda} = \bar H_{\mu\nu\lambda}
- \Bigl( V_{\mu} H_{\nu\lambda}
+ cyclic ~permutations \Bigr)
\end{equation}
This is the form which we need to dualize in order to
obtain the third gauge field in five dimensions. Switching to the
Einstein frame and introducing the Hodge dual of the
gauge-invariant five-dimensional three-form $X'_{\mu\nu}$ we
get the relation between it and
$\bar H'_{\mu\nu\lambda}$:
\begin{equation}\label{HX}
\bar H'_{\mu\nu\lambda} = \frac{1}{2}
e^{4\phi_5/3} \sqrt{\hat g_5}
\epsilon_{\mu\nu\lambda\alpha\beta}
\hat X'^{\alpha\beta} - \Bigl( V_{\mu} H'_{\nu\lambda}
+ cyclic ~permutations \Bigr)
\end{equation}
Equating the RHS of this equation with the RHS of the second
equation of (\ref{Ddaxred2}), substituting in
$H'_{\mu\nu} = X_{\mu\nu}$ and
recalling again the definition of the dual gauge
field $X$ (\ref{XKRd}),
we finally obtain
\begin{equation}\label{permdu}
X'_{\mu\nu} = H_{\mu\nu} ~~~~~~~~~~~ H'_{\mu\nu} = X_{\mu\nu}
\end{equation}
i.e. exactly the permutation rule we have discussed before.
Thus we see that the six-dimensional string/string
duality map reduces to one of the $D_4$ transformations.

The two transformations discussed above are
sufficient to describe the full action of the group $D_4$,
because it has only two independent
generators. The result of their combinations,
however, is non-trivial because in combination with
the arbitrariness of the direction taken to reduce
the action from six to five, and further to
four dimensions, they lead to different compactifications to four
dimensions, interchanging four-dimensional real moduli with the
complex axidilaton, which results in exchanges between the
$S$ and $T$ duality symmetries. In the approach presented here,
after applying one of these transformations,
one only needs to replace the new $X$ field with its
dual three-form
and complete the reduction of the model to four dimensions,
reading off the relevant degrees of freedom.

As an interesting consequence of this symmetry
there appears a triadic relationship among the fundamental string,
the solitonic string and a singular Brinkmann pp wave
in six dimensions. As these solutions all possess the same
number of supersymmetries in six dimensions,
in the theory in which they are embedded,
perhaps this triangular
relationship should not come as too much of a surprise.
Let us now outline how the transformations we have
discussed above establish a connection between these solutions.
The fundamental string solution is \cite{dhgr}
\begin{eqnarray}\label{fund}
d\bar s^2 &=&\Bigl( 1 + \frac{C}{r^2}\Bigr)^{-1}
(-dt^2 + dy^2) + dr^2 + r^2 d\l_3
\nonumber \\
\bar B &=& \frac{C}{C + r^2} dy \wedge dt \\
e^{- \Phi} &=& 1 + \frac{C}{r^2} \nonumber
\end{eqnarray}
with $dl_3$ the line element of the unit three-sphere $S^3$.
It represents the target space configuration of
an elementary string
state in six dimensions, and is singular at $r=0$.
The solitonic string solution can be obtained from
(\ref{fund}) by applying the
duality map defined by the equation (\ref{Duff2}),
or equivalently, by reducing on $y$,
applying the triangle transformation $\sigma'=(\sigma-\phi_5)/2$,
$\phi'_5=-(3\sigma+\phi_5)/2$, $B \leftrightarrow X$,
and then lifting the new configuration
back to six dimensions. The result is \cite{dk,duff2,duffetal}
\begin{eqnarray}\label{solit}
d\bar s^2 &=& -dt^2 + dy^2
+ \Bigl( 1 + \frac{C}{r^2} \Bigr) dr^2
+ \Bigl( C + r^2 \Bigr)  d\l_3
\nonumber \\
\bar H &=& - 2C \omega_3  \\
e^{- \Phi} &=&
\Bigl(1 + \frac{C}{r^2}\Bigr)^{-1} \nonumber
\end{eqnarray}
The $\omega_3$ is the volume form
of the unit three-sphere $S^3$. This solution is actually
nonsingular as $r \rightarrow \infty$, as can be seen by
changing coordinates to $\rho = \ln r$, and realizing that
$\rho \rightarrow - \infty$
is geodesically excluded from the manifold.
This, and the fact that the string axion charge
is carried not by matter sources but by the topology
of the target space, is the reason
why the solution is called solitonic.

In the above two solutions, the string charge was carried
by the axion field. As we have seen above, from the
triangular symmetry there still exists the possibility to
trade the axionic charge for the Kaluza-Klein one.
Applying the five-dimensional scale factor
duality $\sigma \rightarrow - \sigma$, $V \leftrightarrow B$
to the fundamental string
(\ref{fund}), we find
\begin{equation}\label{brink}
d\bar s^2 = - \Bigl( 1 + \frac{C}{r^2} \Bigr)^{-1} dt^2 +
\Bigl( 1 + \frac{C}{r^2} \Bigr)
\Bigl(dy - \frac{C}{C + r^2} dt \Bigr)^2 + dr^2 + r^2 d\l_3
{}~~~~ \bar B = 0 ~~~~ \Phi = 0
\end{equation}
Note that the matter fields are trivial. Changing the
coordinates to $u=t-y$, $v=t+y$, we
can rewrite the solution as
\begin{equation}\label{brink1}
d\bar s^2 = dudv + \frac{C}{r^2} du^2 + dr^2 + r^2 d\l_3
{}~~~~~~ \bar B = 0 ~~~~~~  \Phi = 0
\end{equation}
which we immediately recognize as a Brinkmann pp wave
singular at $r=0$ \cite{tsey,hortse,kall}. Strictly speaking, this
metric has an undesirable property that if $y$ is compactified
periodically there appear closed null curves, manifest after the
transformation to null coordinates.
Also, from the string point of view,
this solution is infinitely degenerate,
because we can absorb $C$ away by a coordinate
transformation $du \rightarrow du' = du \sqrt{C}$,
$dv \rightarrow dv' = dv /\sqrt{C}$. It is then not
difficult to see that the last $D_4$ inversion, characterized
by $\sigma'=(\sigma+\phi_5)/2$, $\phi'_5=-(3\sigma-\phi_5)/2$,
$V \leftrightarrow X$ merely interchanges the solitonic
(\ref{solit}) and the pp wave (\ref{brink1}) solutions.
Thus there are no other non-trivial $D_4$ images of the fundamental
string (\ref{fund}).

There are two immediate observations based on the triangular
correspondence of these string solutions. First, we know that
both the fundamental string and the pp wave are exact solutions to
all orders in $\alpha'$. This comes about because both
solutions have a conserved chiral current on the
world-sheet (being an $F$-type (fundamental) and a $K$-type
conformal field theory (pp wave) discussed
recently by Horowitz and Tseytlin \cite{hortse}). This establishes
that the $T$-duality
relationship between the pp wave and the fundamental string
is in fact exact, modulo field redefinitions. Namely, the
string subtraction schemes in which these two solutions are exact
may differ by finite renormalizations (i.e., string
field redefinitions), implying that the $T$ duality could also
pick up these corrections.
Given this and the triangular relationship
of these solutions with the solitonic string, it is then compelling
to conjecture that there should exist a scheme in which the
solitonic string is also exact to all orders in $\alpha'$. This
would be necessary to establish the equivalence of all the $D_4$
transformations. An argument
regarding the exactness of the solitonic string was put forth
earlier, but it required the introduction of a Yang-Mills gauge
field to cancel $\alpha'$ corrections \cite{dk}. In light of the
triangular relationship, however, this may not be necessary.
A related observation is that since both the fundamental and solitonic
strings are identified as the states in string spectrum, for $D_4$
to be an exact symmetry of string theory the pp wave must also belong
to this spectrum. Since the wave is infinitely degenerate, and
it has $S^3$ target-space symmetry it would be interesting to see
if it can be related to a five-dimensional variant of
the recently conjectured massless black holes associated
with conifold singularities \cite{strom}.

\vskip 1.0truecm

{\it Note added.} Upon the completion of the research presented here,
I have received
the paper entitled ``Four Dimensional String/String/String Triality"
by M.J. Duff, J.T. Liu and J. Rahmfeld,
preprint \# CTP-TAMU-27/95/hep-th/9508094, \cite{trial}
where the triality relationship among the heterotic, type $II_A$
and type $II_B$ string theory was established, and which overlaps
in some length with the present work. Also, a related analysis
is presented in the paper
``U-Dualtiy and Integral Structures"
by P.S. Aspinwall and D.R. Morrison,
preprint \# CLNS-95/1334/hep-th/9505025, \cite{am}, since
published in Phys. Lett. {\bf B355} 141.

\vskip 1.0truecm
\noindent {\bf Acknowledgements}
\vskip 1.0truecm
The author would like to thank R.R. Khuri and
R.C. Myers for stimulating conversation and helpful
comments on the manuscript.
This work was supported in part by NSERC
of Canada, and in part by an NSERC postdoctoral
fellowship.

\end{document}